\documentclass[10pt, conference]{IEEEtran}
\usepackage[caption=false]{subfig}
\usepackage{algorithm}
\usepackage{algpseudocode}
\usepackage{adjustbox}
\usepackage{amsmath,amssymb,amsfonts}
\usepackage{graphicx}
\usepackage{url} 
\usepackage{xcolor}
\def\BibTeX{{\rm B\kern-.05em{\sc i\kern-.025em b}\kern-.08em
    T\kern-.1667em\lower.7ex\hbox{E}\kern-.125emX}}

\begin{document}
\title{COUNTER: Cluster GCN based Energy Efficient Resource Management for Sustainable Cloud Computing Environments}

\author{\IEEEauthorblockN{1\textsuperscript{st} Han Wang}
\IEEEauthorblockA{\textit{School of Electronic Engineering and Computer Science} \\
\textit{Queen Mary University of London}\\
Mile End Road, London, United Kingdom \\
hanwang@qmul.ac.uk}
\and
\IEEEauthorblockN{2\textsuperscript{nd} Sukhpal Singh Gill}
\IEEEauthorblockA{\textit{School of Electronic Engineering and Computer Science} \\
\textit{Queen Mary University of London}\\
Mile End Road, London, United Kingdom \\
s.s.gill@qmul.ac.uk}
\and
\IEEEauthorblockN{3\textsuperscript{rd} Steve Uhlig}
\IEEEauthorblockA{\textit{School of Electronic Engineering and Computer Science} \\
\textit{Queen Mary University of London}\\
Mile End Road, London, United Kingdom \\
steve.uhlig@qmul.ac.uk}}

\maketitle
 
\begin{abstract}
Cloud computing, thanks to the pervasiveness of information technologies, provides a foundational environment for developing IT applications, offering organizations virtually unlimited and flexible computing resources on a pay-per-use basis. However, the large data centres where cloud computing services are hosted consume significant amounts of electricity annually due to Information and Communication Technology (ICT) components. This issue is exacerbated by the increasing deployment of large artificial intelligence (AI) models, which often rely on distributed data centres, thereby significantly impacting the global environment. This study proposes the COUNTER model, designed for sustainable cloud resource management. COUNTER is integrated with cluster graph neural networks and evaluated in a simulated cloud environment, aiming to reduce energy consumption while maintaining quality of service parameters. Experimental results demonstrate improvements in resource utilisation, energy consumption, and cost effectiveness compared to the baseline model, HUNTER, which employs a gated graph neural network aimed at achieving carbon neutrality in cloud computing for modern ICT systems.
\end{abstract}


\section{Introduction}
The ongoing digitisation has elevated computing resources to the status of an essential utility in today's digital world, contributing significantly to the widespread adoption of cloud computing \cite{StockAICloud}. Cloud computing provides flexible, on-demand, and highly customisable cloud services \cite{singh2017journey}, supplying virtually unlimited resources and supporting developments across various sectors of society \cite{zhou2020energy}. Organisations across different industries benefit from reliable computing resources, enabling them to focus on their core operations and reduce costs by avoiding investments in physical IT hardware \cite{xu2022coscal}. However, the extensive application of cloud computing services has substantial environmental implications, as large Cloud Data Centres (CDCs) typically consume a considerable amount of energy due to contemporary Information and Communication Technology (ICT) equipment \cite{AmazonAICloud}. According to the International Energy Agency, CDCs and Artificial Intelligence (AI) consumed approximately 460 TWh of electricity worldwide in 2022, accounting for roughly 2\% of global electricity consumption \cite{2024Energy} and this consumption is projected to double by 2026. In addition to energy consumption, data centres in the United States consumed over 600 billion litres of water in 2014, with some cloud service providers  sourcing more than half of their water from potable supplies \cite{Mytton2021}. Consequently, regulatory bodies in the EU and the US are continuously formulating new guidelines to improve the energy efficiency of cloud providers and minimise their environmental footprint \cite{2024Energy}. Therefore, it is essential to develop more sustainable approaches to ensure the continued growth and viability of cloud computing.

\subsection{Motivation and Contributions}
Recent advancements in AI have significantly expanded its applications and accessibility to the public, creating transformative impacts across various sectors of society \cite{HealthAIoT}. For instance, generative AI models such as ChatGPT have notably influenced education by assisting language learning, computer programming, and content creation,  thus enhancing productivity \cite{GAIKube}. Similarly, DeepMind’s AlphaFold  has demonstrated AI’s potential for solving complex scientific problems, such as protein structure prediction \cite{Haque2024}.

Despite increasing scientific evidence of AI’s positive social impact, substantial research highlights its environmental consequences  \cite{HealthEdgeAI}. The carbon footprint associated with cloud computing and AI cannot be overlooked \cite{Gupta2021, Wang2024}. Both the training and inference phases of AI models demand tremendous computational power, which is primarily provided by large, distributed data centres \cite{AINet0}. The dependence of AI advancements on cloud services has thus resulted in increased energy consumption and water usage. Generally, larger and more advanced models lead to greater carbon emissions, predominantly due to the high power demands of computing processors and cooling systems \cite{Gaur2023, Shaji2023}. The substantial energy use and resulting carbon emissions underline the necessity for sustainable cloud computing solutions, not only to address global climate challenges but also to ensure the long-term viability of cloud computing itself, as reducing energy consumption additionally lowers the costs associated with expensive AI model training, contributing to enhanced sustainability \cite{McDonald2022}. Furthermore, various cloud architectures and operational strategies can significantly influence the energy consumption of CDCs, emphasising the importance of optimising these operational strategies to effectively reduce energy usage \cite{Khomh2018, AmazonAICloud}. The main contributions of this paper are:
\begin{itemize}
    \item A resource management model, COUNTER is proposed to optimise cloud resource utilisation and reduce energy consumption within cloud environments.
    \item The performance of COUNTER model that employs cluster graph neural network is evaluated and compared with baseline model HUNTER.
    \item The effectiveness of cluster graph neural network algorithms is evaluated in environment with complex relational structures.
\end{itemize}

The rest of this paper is organised as follows: Section II reviews existing literature on cloud resource management. Section III details the proposed methodology and describes the architecture of the COUNTER model. In Section IV, the experimental setup is presented, and the performance of the proposed model is analysed through comparison with the baseline model. Section V summarises the key findings and discusses directions for future research.

\section{Related Work}
Saxena et al. \cite{Saxena2023} proposed a secure and sustainable load management model based on a multi-layered feed-forward neural network to improve resource utilisation, reduce energy consumption, and increase power usage efficiency. They simulated the proposed model using Google Cluster dataset on physical server machines. The results demonstrated improvements in workload prediction, resource utilisation, and power consumption compared to existing methods. However, their primary contributions focused more on predicting oncoming workloads rather than optimising individual resource components. Gill et al. \cite{CRUZE} developed CRUZE, a holistic resource scheduling framework that employs the cuckoo optimisation algorithm to manage resources such as servers, network components, storage, and cooling systems. Experimental results conducted on CloudSim indicated that the proposed technique reduced energy consumption by 20.1\%, achieved through enhanced resource utilisation and energy efficiency. Xu et al. \cite{Xu2022EsDNN} developed a framework integrating efficient deep neural networks with gated mechanisms to predict cloud workloads, using datasets from Google Cloud Platform and Alibaba Cloud. Although the approach demonstrated improved workload prediction, their research did not explore its potential for optimising energy consumption. 

Sun et al. \cite{Sun2024} proposed the Sgp-Stream framework to optimise system elasticity and resource scheduling in large-scale data streaming scenarios, such as big data streaming and real-time processing. Their framework was implemented on Apache Storm with a real-world dataset from Twitter, their experiments showed significant reductions in latency and notable improvements in throughput and resource utilisation. Kumar et al. \cite{Kumar2021} proposed an approach that combined the blackhole optimisation algorithm with a self-directed neural network to improve workload prediction. They validated their model using various datasets, including traces from NASA Web Server Logs, Google Cloud Platform, and several university web servers, achieving significant enhancements in predictive accuracy. Tuli et al. \cite{HUNTER} developed HUNTER, a holistic cloud resource management model based on gated graph convolutional networks. Their approach integrated factors such as CPU and RAM utilisation, SLA violations, execution time, and thermal characteristics. Evaluations performed using CloudSim and COSCO on Microsoft Azure demonstrated superior Quality of Service (QoS) performance and reduced energy consumption compared to prior approaches. This model serves as the baseline for comparison in this research.

\subsection{Critical Analysis}
Table \ref{comparison_table} summarises recent research in cloud computing resource management, highlighting various approaches to cloud resource optimisation. The comparison considers experimental environments, optimisation techniques, datasets used, and key contributions. The reviewed studies incorporate both traditional heuristic optimisation techniques and machine learning algorithms, primarily targeting improvements in workload prediction accuracy, resource utilisation efficiency, and energy optimisation. Their primary optimisation objectives include: 

\begin{itemize}
    \item Cloud Workload Prediction: Predicting future cloud resource demands to enable proactive and efficient resource allocation.
\end{itemize}
\begin{itemize}
    \item Resource Optimisation: Improving the utilisation of resources such as CPU, GPU, and RAM to enhance overall system efficiency.
\end{itemize}
\begin{itemize}
    \item Cloud Energy Optimisation: Minimising energy consumption of CDCs through strategies such as cooling system adjustments and power supply management optimisation.
\end{itemize}

\begin{table*}
\caption{Comparison of the proposed COUNTER framework with existing works.}
\label{comparison_table}
\begin{center}
\resizebox{\textwidth}{!}{
\begin{tabular}{|c|c|c|c|c|}
\hline
\textbf{Work} & \textbf{Tool} & \textbf{Optimisation Technique}  & \textbf{Dataset} & \textbf{Key Contributions} \\ \hline

\cite{Saxena2023} & Physical Machine & Neural Network & Google Cluster Dataset & Workload Prediction \\ \hline

\cite{CRUZE} & CloudSim & Cuckoo Algorithm & Heterogeneous Workloads & Resource and Energy Efficiency \\ \hline

\cite{Xu2022EsDNN} & TensorFlow & esDNN with Gated Mechanism & Alibaba Cloud and Google Cluster Dataset & Workload Prediction \\ \hline 

\cite{Sun2024} & Apache Storm &Sgp-Stream Framework & Twitter Dataset & Resource Optimisation and Scalability \\ \hline 

\cite{Kumar2021} & Matlab & Self-Directed Learning with Blackhole Algorithm & NASA, Google Cloud, Web Servers Dataset & Workload Prediction\\\hline 

\cite{HUNTER} & CloudSim, COSCO on Microsoft Azure & Gated Graph Convolution Network & Defog & Resource and Energy Optimisation \\ \hline

\textbf{COUNTER (This Work)} & Philharmonic & Cluster Graph Neural Network & Bitbrains & Resource and Energy Optimisation and Cost \\ \hline
\end{tabular}}
\end{center}
\end{table*}

\begin{figure}[ht]
    \centering
    \includegraphics[width=1\linewidth]{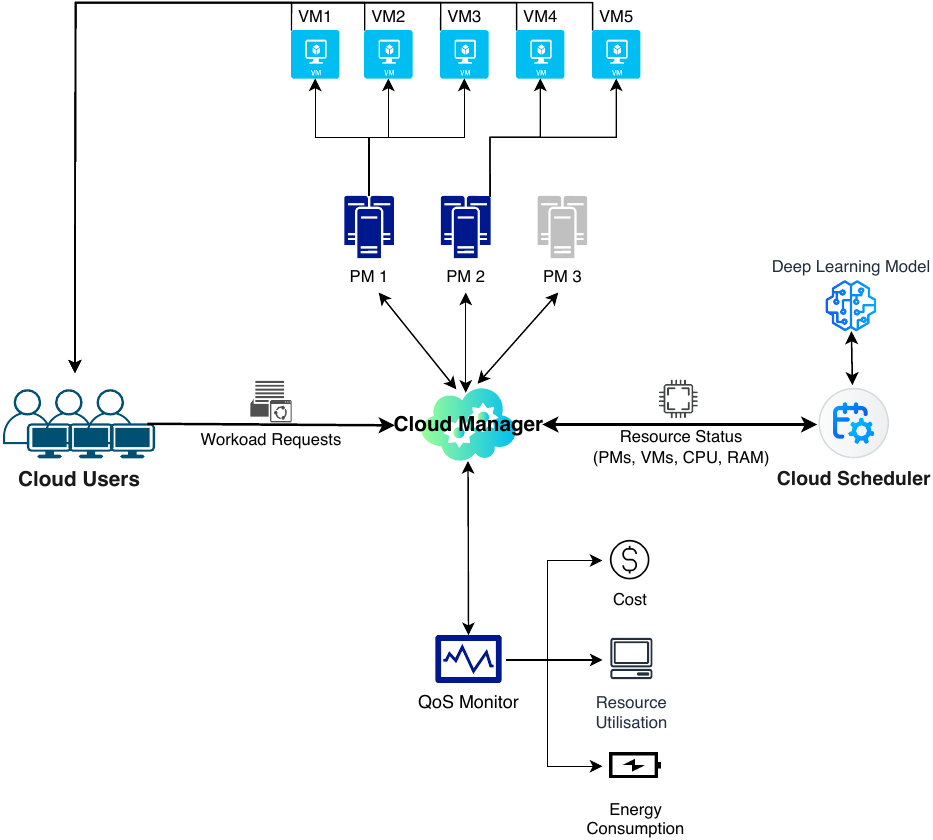}
    \caption{COUNTER Architecture}
    \label{fig:Cloud Architecture}
\end{figure}

\section{METHODOLOGY}
This section explains the proposed COUNTER model, describes its architecture, and presents a diagram illustrating its main components.

\subsection{COUNTER: Cluster GCN-Based Resource Management for Sustainable Cloud Computing Environments}

The primary contribution of this research is the development of a sustainable cloud resource management model named COUNTER. The objective of COUNTER is to manage cloud resources efficiently to reduce energy consumption and improve the sustainability of cloud computing environments. This focus is especially critical given that major CDCs components, such as processors, memory, networking, storage, and cooling systems, consume significant amounts of energy, with processors often being the largest contributor to overall energy usage. The COUNTER model accounts for these key cloud components, as illustrated in Fig.~\ref{fig:Cloud Architecture}. Its structure consists of the following main elements:

\begin{itemize}

 \item \textbf{Cloud Users}: Cloud users initiate workload requests within the cloud environment. Each request specifies resource requirements such as CPU frequency, number of CPU cores, RAM capacity, and the expected duration of resource usage. These parameters collectively define the resources needed for executing the workload. This structured information is extracted from the cloud workload dataset.
 
 \item \textbf{Cloud Manager}: The cloud manager oversees the entire cloud environment. It abstracts the underlying physical infrastructure, manages interactions with cloud users, and supervises the lifecycle of virtual machines (VMs), including operations such as booting, pausing, migrating, and deleting VMs. Additionally, it calculates QoS parameters for the overall system.
 
 \item \textbf{Cloud Scheduler}: Cloud scheduler is integrated with a machine learning model, determining optimal scheduling decisions, allocating VMs to appropriate physical machines (PMs) based on resource availability and user requests. The scheduler is central to workload distribution and resource balancing across available PMs. In this research, a cluster graph neural network is employed to develop the scheduler, aiming to optimise resource utilisation and minimise energy consumption.
 
\end{itemize}

The COUNTER model is built upon a Clustering Graph Neural Network (Cluster GCN). Cluster GCN is a recent advancement in graph-based neural network methods, employing graph clustering techniques to handle large-scale graph data efficiently. It enhances training efficiency by creating mini-batches from clustered graph data to achieve greater scalability as the size of the graph increases \cite{Chiang2019}. This allows for efficient training on complex graph structures without incurring excessive computational costs, thereby enhancing accuracy and generalisation capability.

The pseudocode for the COUNTER model is shown in Algorithm~\ref{Pseudocode}. Workload requests ${W_i}$ are fetched from cloud users to cloud manager, the cloud manager collects current resource availability information {$R_i$} from PMs. Both workload requests and resource availability information are forwarded to the cloud scheduler, which then generates scheduling decisions based on this information. Finally, the cloud manager executes these scheduling decisions, allocating workloads to suitable PMs, updating their resource status, and computing relevant QoS metrics.

\begin{algorithm}[ht]
\caption{COUNTER Model}
\label{Pseudocode}
\scriptsize
\begin{algorithmic}[1]
\State \textbf{Input:} Workload requests ${W_i}$ from Cloud Users
\State \textbf{Output:} Optimal Assignment of VMs to PMs
\Procedure{COUNTER}{$W_i$}
    \Procedure{CloudManagerCollect}{$W_i$}
        \ForAll{$R_i$ in Workload Requests}
            \State Retrieve workload request {$W_i$} from Cloud Users
            \State Collect current resource availability {$R_i$} from PMs
            \State Send {$R_i$} and {$W_i$} to Cloud Scheduler
            \State $Schedule \gets$ \Call{CloudScheduler}{$R_i$, $W_i$}
        \EndFor        
        \State \Return $Schedule$
    \EndProcedure
    \Procedure{CloudManagerExecute}{$Schedule$}
        \State Execute VM allocation actions based on schedule
        \State Update PM resource statuses
        \State Calculate QoS parameters
    \EndProcedure
\EndProcedure
\end{algorithmic}
\end{algorithm}

\subsection{QoS Parameters}
In cloud computing research, the priority and significance of QoS parameters vary depending on the specific research objectives. For instance, studies aimed at optimising cloud load balancing often prioritise QoS parameters such as response time, migration time, fault tolerance, scalability, and load imbalance, aiming to improve system throughput and overall performance \cite{Kumar2019}. In contrast, sustainable cloud computing research typically emphasises QoS parameters such as energy efficiency and resource utilisation \cite{Saxena2023, 2021SustainableAI}. In this study, the COUNTER model optimises QoS parameters from the perspective of cloud service providers, specifically focusing on resource utilisation, energy consumption, and operational cost.

\section{PERFORMANCE EVALUATION}
This section details the experimental configurations and simulation environment, followed by an analysis of the obtained experimental results.

\subsection{Experimental Setup}
In this research, the proposed COUNTER model considers various input parameters within a simulation environment to optimise QoS parameters of resource utilisation, energy consumption and cost. The model was conducted using the Philharmonic cloud simulation framework \cite{Lucanin2016, Lucanin2014}. The experiment calculates energy consumption costs based on a dynamic electricity pricing dataset integrated within the simulator. The input parameters, including simulation duration, number of PMs, number of VMs,  and the CPU and RAM capacities are configured as detailed in Table~\ref{tab:Configuration Settings}.

\subsection{Workloads}
The GWA-T-12 cloud workload dataset from Bitbrains is used in this research. This dataset contains VM workload traces formatted as time series data, originating from services operating across various sectors such as banking and insurance \cite{Shen2015, Bitbrains}. Workloads within the simulation environment are generated based on the specified simulator configurations combined with the Bitbrains dataset.

\begin{table}[ht]
\caption{Simulation Configuration Settings}\label{tab:Configuration Settings}
\centering
\resizebox{0.35\textwidth}{!}{
\begin{tabular}{cc}
\hline
\textbf{Configuration Types} & \textbf{Specifications} \\ \hline
        Number of Physical Machines & 8 \\
        CPU Cores per Server & 32 \\
        Minimum CPU Frequency & 1600 MHz \\
        Maximum CPU Frequency & 3400 MHz \\
        RAM Capacity Per Server & 16 GB - 64 GB \\
        Number of Virtual Machines & 60 \\
        Minimum VM Runtime & 1 hour \\
        Maximum VM Runtime & 48 hours \\
        Total Simulation Duration & 120 hours \\
        Peak Power per Physical Machine & 200 Watts \\
        Idle Power of Physical Machine & 100 Watts \\
\hline
\end{tabular}}
\end{table}

\subsection{Baseline Model}
This section explains the baseline model HUNTER and compares it with the proposed COUNTER model. Both models employ different neural network architectures: the HUNTER model uses a gated graph convolutional recurrent neural network, whereas COUNTER uses a cluster graph neural network. The gated graph convolutional recurrent neural network integrates a gating mechanism to control information flow within graph neural networks \cite{2022Theory, Ruiz2020}. The gating mechanism dynamically and automatically determines which information should be propagated through the network and how much previous information should be retained, depending on the context of the data. This significantly enhances the traditional graph neural network’s ability to manage temporal dynamics in complex systems \cite{Wang2022}.

\subsection{Results}

\begin{figure}[htbp]
    \centering
    \subfloat[Energy Consumption Comparison]{%
        \includegraphics[width=0.5\linewidth]{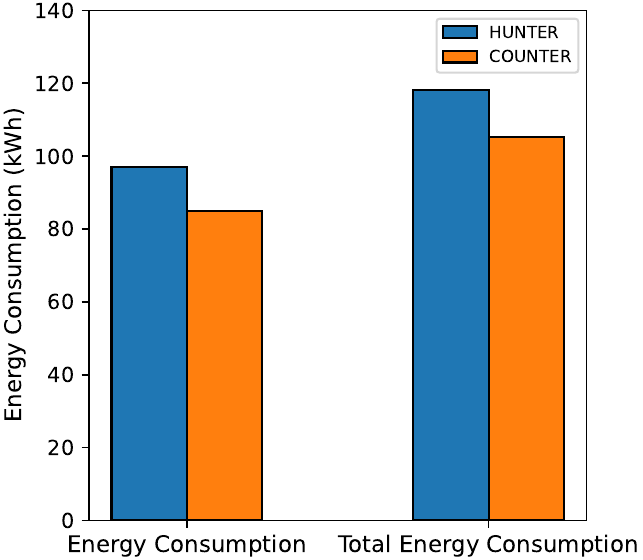}%
        \label{fig:EnergyConsumption}%
    }\hfill
    \subfloat[Total Energy Cost Comparison]{%
        \includegraphics[width=0.5\linewidth]{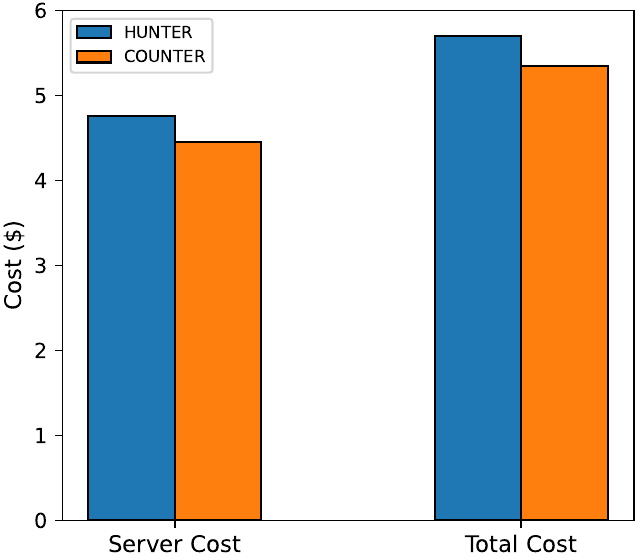}%
        \label{fig:EnergyCost}%
    }
    \caption{Performance Comparison of COUNTER and HUNTER in terms of Energy Consumption and Cost}
    \label{fig:Comparison}
\end{figure}

\begin{figure}[ht]
    \centering \includegraphics[width=1\linewidth]{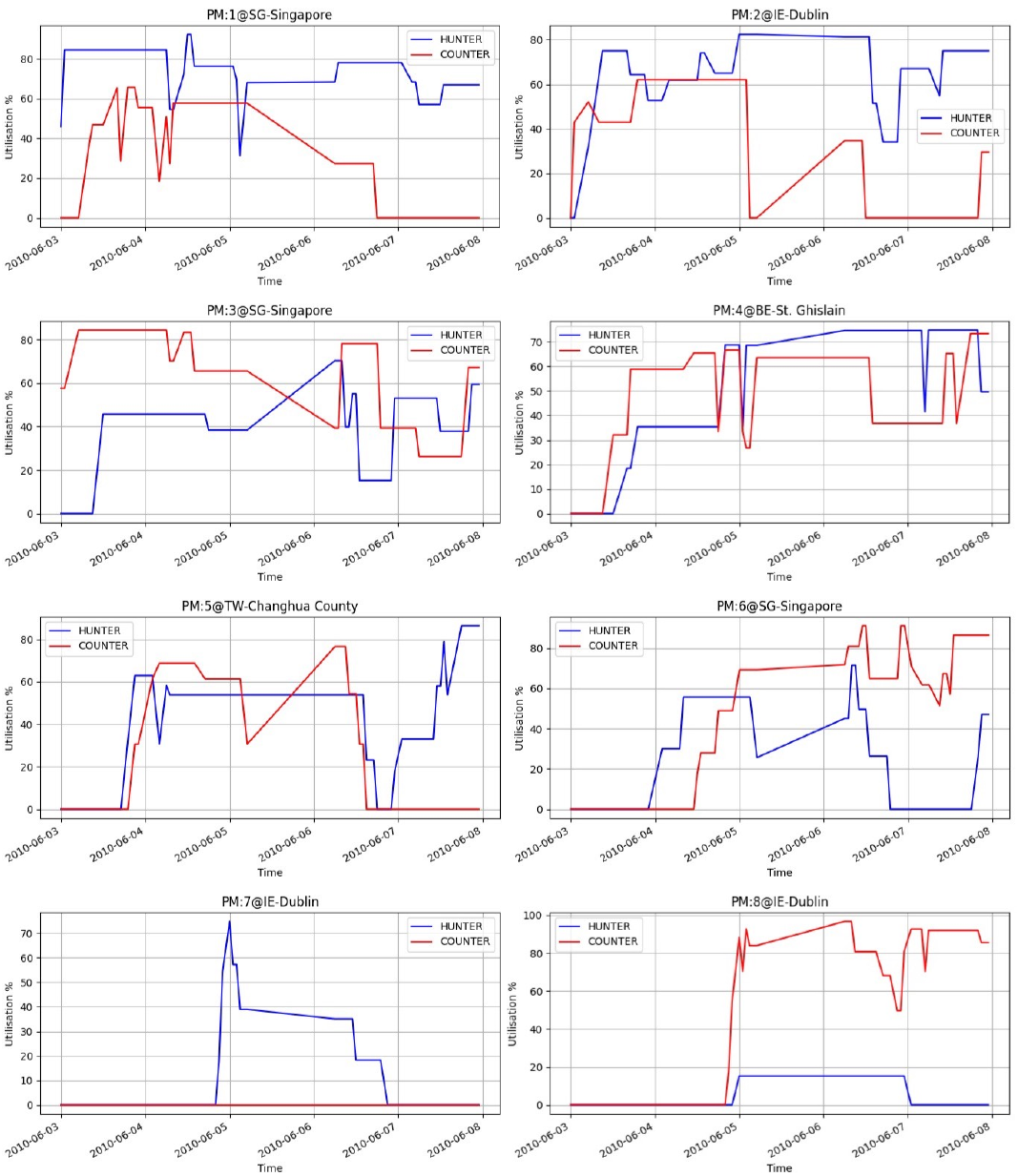}
    \caption{Performance Comparison of COUNTER and HUNTER in terms of PMs Resource Utilisation}
    \label{fig:PMs resource}
\end{figure} 

In the simulation experiment, three QoS parameters including resource utilisation, energy consumption and energy cost are compared between the baseline model HUNTER and the proposed COUNTER model under identical configuration settings. Fig.\ref{fig:PMs resource} shows the resource utilisation rates of eight PMs at different locations during the simulation period. Results indicate that the COUNTER model achieves a higher maximum resource utilisation rate of 96.7\% compared to HUNTER of 92.1\%. The resource scheduling strategy of COUNTER focuses on increasing the utilisation rate of each PM while minimising the number of concurrently active servers. Consequently, Fig. \ref{fig:PMs resource} demonstrates that COUNTER maintains fewer active servers simultaneously compared to HUNTER. Fig. \ref{fig:Comparison} shows the energy consumption and energy cost comparisons for the two models. The total energy consumption \textit{(E)} used in this study, including cooling systems, is calculated as follows \cite{Li2018}:
\begin{equation}
    E = E_{\text{Processor}} + E_{\text{Cooling}} + E_{\text{Extra}}
    \label{eq:energy_consumption}
\end{equation}

Where \( E \) represents total energy consumption; \( E_{\text{Processor}} \) represents energy consumed by the processor; \( E_{\text{Cooling}} \) represents energy consumed by the cooling system; \( E_{\text{Extra}} \) represents additional energy use, such as switches and lighting.

The total energy consumption  of the COUNTER model, including the cooling system, is 105.2KWh compared to 118.2KWh for HUNTER model. Consequently, COUNTER achieved lower energy costs than HUNTER. Experimental results demonstrated that COUNTER provided more effective resource scheduling, lower energy consumption, and improved QoS outcomes. Further experiments will be conducted using diverse datasets to enhance the generalisability and performance of the proposed model.

\section{CONCLUSIONS AND FUTURE WORK}
Cloud computing services plays a significant role in the contemporary technological landscape. However, their widespread adoption across numerous sectors has raised considerable concerns regarding environmental impacts. To address these concerns, this study proposed COUNTER, a sustainable cloud resource management model designed to minimise the environmental impact of cloud computing by enhancing resource utilisation efficiency, thereby reducing energy consumption and contributing towards carbon-neutral computing. The COUNTER model integrates a cluster GCN and was implemented using the Philharmonic simulation environment. Experimental results indicated that COUNTER outperforms the baseline HUNTER model, achieving notable improvements in resource utilisation, energy consumption, and overall operational costs. The findings demonstrated the effectiveness of cluster GCN in managing complex relational structures typical of cloud computing environments.

In future work, the proposed COUNTER model can be deployed and evaluated in real-world cloud environments, such as Google Cloud Platform and Microsoft Azure, to further validate its practical performance. Additionally, future studies may incorporate other advanced machine learning algorithms and utilise more diverse datasets to improve the model’s efficiency and generalisability across different infrastructure scenarios. Further experiments should consider additional cloud environment parameters, such as the energy consumption associated with networking and memory resources, which could lead to more comprehensive resource scheduling optimisation.

\section*{Software Availability}
The code for the proposed COUNTER framework is available at \url{https://github.com/HTXW/COUNTER}

\bibliography{reference}
\bibliographystyle{IEEEtran}
\end{document}